\documentclass[sigplan, screen]{acmart}
\AtBeginDocument{%
  \providecommand\BibTeX{{%
    \normalfont B\kern-0.5em{\scshape i\kern-0.25em b}\kern-0.8em\TeX}}}

\usepackage[ruled,vlined,linesnumbered]{algorithm2e}

\SetCommentSty{mycommfont}

\usepackage{xcolor}
\usepackage{color}
\definecolor{codegreen}{rgb}{0,0.6,0}
\definecolor{codegray}{rgb}{0.5,0.5,0.5}
\definecolor{codepurple}{rgb}{0.58,0,0.82}
\definecolor{backcolour}{rgb}{0.95,0.95,0.92}
\definecolor{textblue}{rgb}{.2,.2,.7}
\definecolor{textred}{rgb}{0.54,0,0}
\definecolor{textgreen}{rgb}{0,0.43,0}
\definecolor{codered}{rgb}{201,72,12}

\usepackage[T1]{fontenc}
\usepackage[scaled=0.85]{beramono} 
\usepackage{listings}
\usepackage{multirow}

\lstdefinestyle{tt1}{
language=C,
basicstyle=\linespread{0.9}\ttfamily\footnotesize,
breaklines=true,
numbers=left,
frame=single,
numberstyle=\tiny, 
stepnumber=1,
numbersep=5pt, 
tabsize=4,
keywordstyle=\bfseries\color{codegreen},
commentstyle=\color{textred},   
stringstyle=\color{textgreen},
columns=fullflexible,
keepspaces=true,
xleftmargin=\parindent,
showstringspaces=false,
otherkeywords = {True, False},
keywordstyle=[2]\color{codepurple}\bfseries,
keywords=[2]{wmma, b1},
keywordstyle=[3]\color{textblue}\bfseries,
keywords=[3]{fragment, load_matrix_sync, mma_sync, store_matrix_sync},
keywordstyle=[4]\color{codegreen}\bfseries,
keywords=[4]{matrix_a, row_major, column_major, mem_row_major, tf32},
}

\lstdefinestyle{tt3}{
language=C,
basicstyle=\linespread{0.9}\ttfamily\footnotesize,
breaklines=true,
numbers=left,
frame=single,
numberstyle=\tiny, 
stepnumber=1,
numbersep=5pt, 
tabsize=4,
keywordstyle=\bfseries\color{codegreen},
commentstyle=\color{textred},   
stringstyle=\color{textgreen},
columns=fullflexible,
keepspaces=true,
xleftmargin=\parindent,
showstringspaces=false,
otherkeywords = {True, False},
keywordstyle=[2]\color{codepurple}\bfseries,
keywords=[2]{BLK_H, BLK_W},
keywordstyle=[3]\color{textblue}\bfseries,
keywords=[3]{fragment, load_matrix_sync, mma_sync, store_matrix_sync, __shared__},
keywordstyle=[4]\color{codegreen}\bfseries,
keywords=[4]{matrix_a, row_major, column_major, mem_row_major, tf32},
}
\usepackage{array}
\newcommand\blfootnote[1]{%
  \begingroup
  \renewcommand\thefootnote{}\footnote{#1}%
  \addtocounter{footnote}{-1}%
  \endgroup
}
\copyrightyear{2022}
\acmYear{2022}
\setcopyright{rightsretained}
\acmConference[PPoPP '22]{27th ACM SIGPLAN Symposium on Principles and Practice of Parallel Programming}{April 2--6, 2022}{Seoul, Republic of Korea}
\acmBooktitle{27th ACM SIGPLAN Symposium on Principles and Practice of Parallel Programming (PPoPP '22), April 2--6, 2022, Seoul, Republic of Korea}
\acmDOI{10.1145/3503221.3508408}
\acmISBN{978-1-4503-9204-4/22/02}
\usepackage{multirow}

\newcommand{\hlp}[1]{{\underline{\textbf{\textit{#1}}}}}

\newcommand*{\Mname}{QGTC}

\usepackage{mathtools}

\DeclarePairedDelimiter\floor{\lfloor}{\rfloor}
\usepackage{setspace}
\usepackage{color,soul}
\usepackage{subfig}
\newcommand\aecode[1]{\textcolor{ACMDarkBlue}{#1}}

\begin{document}

\title{QGTC: Accelerating Quantized Graph Neural Networks via GPU Tensor Core}

\keywords{GPU Tensor Core, Quantized Graph Neural Networks, High-performance Computing}

\author[Y. Wang et al.]{Yuke Wang*, Boyuan Feng*, Yufei Ding}
\affiliation{
    \vspace{1pt}
  \institution{\{yuke\_wang, boyuan, yufeiding\}@cs.ucsb.edu}
    \vspace{1pt}
  \institution{University of California, Santa Barbara}
\country{}
 }

\setlength{\textfloatsep}{5pt}

\ccsdesc[500]{Computing methodologies~Neural networks}
\ccsdesc[500]{Computer systems organization~Single instruction, multiple data}

\definecolor{amethyst}{rgb}{0.6, 0.4, 0.8}

\begin{abstract}
  Over the most recent years, quantized graph neural network (QGNN) attracts lots of research and industry attention due to its high robustness and low computation and memory overhead. Unfortunately, the performance gains of QGNN have never been realized on modern GPU platforms.
  To this end, we propose the first Tensor Core (TC) based computing framework, \textbf{QGTC}, to support any-bitwidth computation for QGNNs on GPUs.
  We introduce a novel quantized low-bit arithmetic design based on the low-bit data representation and bit-decomposed computation.
  We craft a novel TC-tailored CUDA kernel design by incorporating 3D-stacked bit compression, zero-tile jumping, and non-zero tile reuse technique to improve the performance systematically.
  We incorporate an effective bandwidth-optimized subgraph packing strategy to maximize the transferring efficiency between CPU host and GPU device.
  We integrate QGTC with Pytorch for better programmability and extensibility.
  Extensive experiments demonstrate that QGTC can achieve evident inference speedup (on average $2.7\times$) compared with the state-of-the-art DGL framework across diverse settings. 
\end{abstract}
\maketitle
\blfootnote{$*$: The first two authors contribute equally.}

\vspace{-25pt}
\section{Introduction}
With the popularity surge of the graph neural networks (GNNs)~\cite{GCNConv, GATConv, SageConv}, research around the full-precision GNNs has been widely studied in terms of its algorithms~\cite{GCNConv, GINConv} and execution performance~\cite{ma2019neugraph, wang2019dgl, pyG} over traditional graph analytical methods, such as Random Walk~\cite{huang2021broader}. 
On the other side, quantized GNN~\cite{feng2020sgquant,tailor2020degree} (QGNN) recently attract lots of attention thanks to its negligible accuracy loss, resilience towards malicious attacks, and significantly lower computations and memory overhead. 
We summarize several key features of GNNs that make them intrinsically suitable for quantization. 
\hlp{First}, the adjacent matrix of GNNs is naturally well-suited for quantization, since we only need to use 0/1 to indicate the existence of edge connections. Thus, using low bits for such information can save both memory and computation. 
\hlp{Second}, the quantization of weight and node embedding can also be beneficial. Because the tiny precision loss in quantization can largely be offset by the node information fusion through the iterative neighbor aggregation process of GNNs. The quantization of floating-point numbers can absorb input perturbations from adversarial attacks.

Despite such great theoretical success of QGNN, the realization of such benefits on high-performance GPUs is still facing tremendous challenges. 
Existing GPU-based GNN frameworks~\cite{pyG,wang2019dgl,GNNAdvisor} are designed and tailored for GPU CUDA cores, which are intrinsically bounded by its peak throughput performance and can only handle the byte-based data types (\textit{e.g.}, \texttt{int32}). 
Although quantized computation can be achieved via pure algorithmic emulation, the actual bit-level performance gains could hardly be harvested, since all underlying arithmetic operations still have to rely on those well-defined data types from CUDA/C++ libraries.
\begin{figure}[t] \small
    \centering
    \includegraphics[width=\columnwidth]{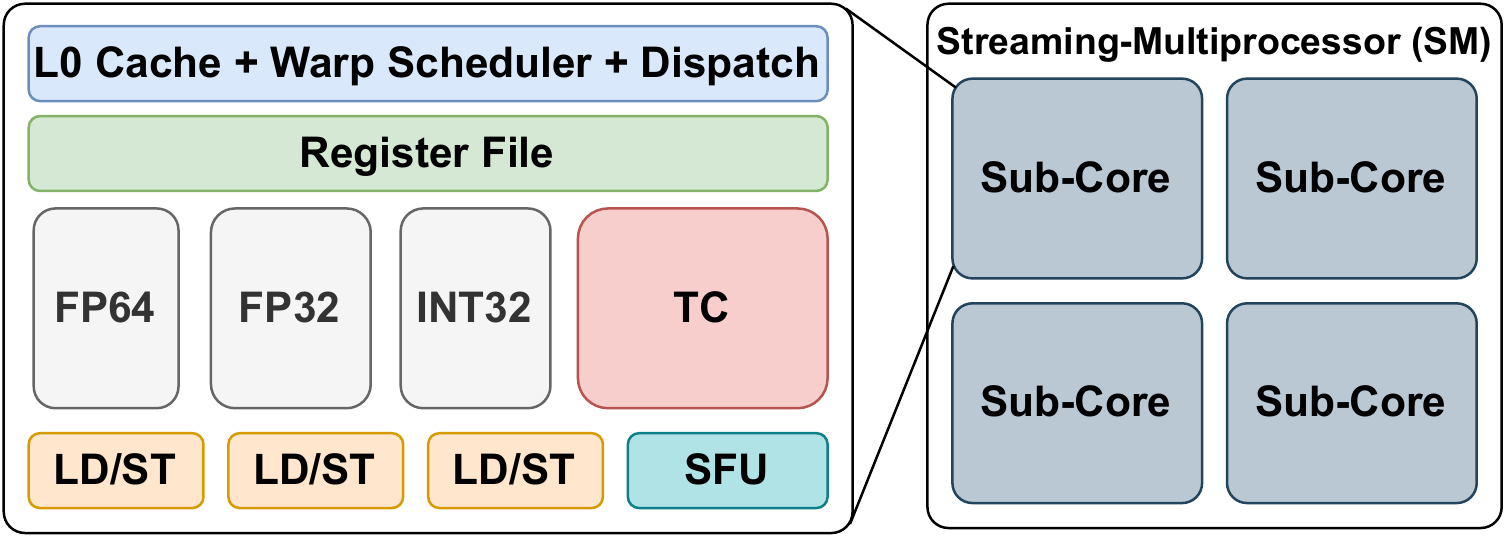}
    \caption{GPU Streaming-Multiprocessor (SM) with TC Design. Note that FP64, FP32, INT32, LD/ST, and SFU are double-precision, single-precision, integer, load/store, and special function units, respectively.}
    \label{fig: SM Hardware Design}
    \vspace{2pt}
\end{figure}

{To tackle these challenges, we decide to move forward with the recent GPU hardware feature -- \textbf{Tensor Core} (TC). The modern NVIDIA GPU with TC design is illustrated in Figure~\ref{fig: SM Hardware Design}. 
TC provides the native support of bit-level operations (\texttt{XOR}, \texttt{AND}), which could be the major ingredient for quantized computation. 
Besides, TC can easily beat CUDA core with a significantly higher throughput performance (more than $10\times$) on conventional NN operations (\textit{e.g.}, linear transformation and convolution). This demonstrates the potential of using TC in accelerating QGNNs.
However, directly using TC for QGNN computation is encountering several challenges. 
\begin{figure}[t] \small
    \centering
    \includegraphics[width=\columnwidth]{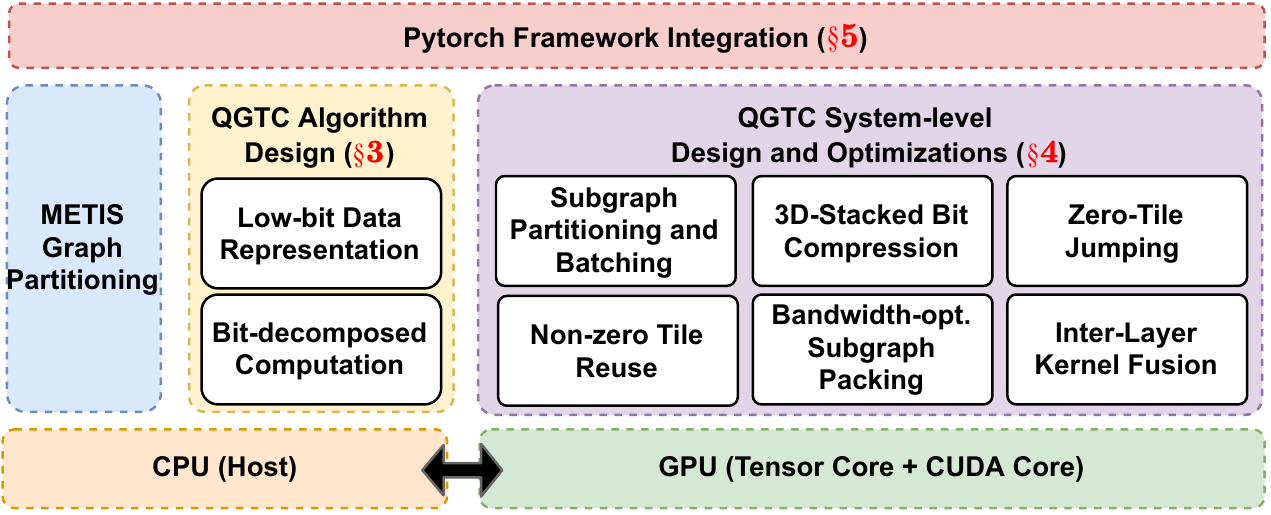}
    \vspace{-15pt}
    \caption{QGTC Overview.}
    \label{fig: QGTC Overview}
\end{figure}
\hlp{First}, the current TC can only support limited choices of bitwidth (\textit{e.g.}, 1-bit and 4-bit), which may not be able to meet the demands of users for any-bitwidth (\textit{e.g.}, 2-bit) computation. 
\hlp{Second}, TC initially tailored for GEMM computation may not directly fit the context of sparse GNN computation. A huge amount of computation and memory access efforts would be wasted on those non-existed edges. This is because the hard constraint of TC input matrix tile-size (\textit{e.g.}, 8$\times$128 for 1-bit GEMM) has to be satisfied, which may require excessive zero paddings. 
\hlp{Third}, the low-bit computation would cause the compatibility issue, since the existing deep-learning frameworks~\cite{PyTorch,tensorflow} cannot directly operate on the low-bit data type.

Therefore, we remark there are several aspects to be considered in order to use TC for QGNNs:}
1) \textbf{\textit{Hardware-level Support}}. 
This inspires us to explore the high-performance GPU hardware features that can efficiently support the QGNN computation. 
Even though it is hard to find such a GPU hardware feature that can directly support any-bitwidth QGNN, some indirect hardware features would potentially be helpful. 
For example, NVIDIA introduced the 1-bit TC-based GEMM on Turing Architecture, which essentially can be used to composite any-bitwidth GEMM.
2) \textbf{\textit{Software-level Optimizations}}. 
This motivates us to optimize the kernel computation according to the characters of QGNN. 
GNN computation is featured with a highly sparse and irregular scheme. It is intrinsically not favorable for the dense GPU computation flow tailored for the traditional NN operators. 
Thus, how to handle such input-level irregularity from the computation and memory perspectives is essential to the performance of QGNN. 
For example, subgraph partitioning~\cite{METIS} based mini-batch GNN computation has been used to increase the computation efficiency without compromising model accuracy performance.
3) \textbf{\textit{Framework-level Integration}}. 
This encourages us to bridge the gap between quantized low-bit implementations and deep-learning frameworks built for full-precision computation. Therefore, our whole system-level design can be seamlessly integrated with the state-of-the-art mainstream NN frameworks to benefit the execution performance and the developing productivity. 

To this end, we introduce \Mname\footnote{QGTC is open-sourced at \url{github.com/YukeWang96/PPoPP22\_QGTC.git}}, the first framework (Figure~\ref{fig: QGTC Overview}) to support any-bitwidth QGNN on GPU TC.

\textbf{At the input level}, we incorporate the METIS~\cite{METIS} graph partitioning to generate a set of dense subgraphs from the highly irregular and sparse input graphs. 
The insight here is that nodes in real-world graphs are likely to form clusters, and such information can be used to benefit the efficiency of GNN computing and model algorithmic performance.

\textbf{At the algorithm level}, we leverage the insight that any-bitwidth QGNN computation can always be decomposed into the 1-bit computation. Each bit in the output can be generated by different combinations of bits from the input. Thus, we use quantized low-bit data representation and bit-decomposed computation base on the ``atomic'' 1-bit type.

\textbf{At the GPU kernel level}, we craft a low-bit computation design tailored for QGNN computation on batched dense subgraphs. We address the key performance bottleneck of the low-bit GNN computing from the memory and computing perspectives. 
Specifically, we use only 1-bit binarized representation for the subgraph adjacent matrix, which is memory efficient for representing the presence/absence of edge connections between nodes. Besides, we use a 3D-stacked bit-compression technique for maintaining quantized low-bit node embedding features and weights.
In addition, we fully exploit the intra-subgraph sparsity through zero-tile skipping and non-zero tile reuse, which can further avoid unnecessary computations and improve the data locality.

\textbf{At the framework level}, we integrate \Mname~with the state-of-the-art Tensor-based PyTorch~\cite{PyTorch} framework. We introduce the new notion of bit-Tensor data type and bit-Tensor computation and warp them up as a new set of PyTorch API extensions. End-users can directly interact with the \Mname~PyTorch APIs to access all functionalities. This largely improves the programmability and extensibility.

Overall, we summarize our key contributions as:
\begin{itemize}
    \item We propose a novel 1-bit composition technique for any-bitwidth arithmetic design ({$\mathbf{\color{amethyst}\S3}$}), which can support QGNN with diverse precision demands.
    \item We introduce a highly efficient implementation of QGNN ({$\mathbf{\color{amethyst}\S4}$}) built on top of the GPU Tensor Core by applying a series of computation optimizations
    (\textit{e.g.}, subgraph partitioning and batching, and zero-tile jumping) 
    and memory optimizations. 
    (\textit{e.g.}, 3D-stacked bit-compression and non-zero tile reuse).
    \item We integrate \Mname~with PyTorch ({$\mathbf{\color{amethyst}\S5}$}) by introducing bit-Tensor data type and bit-Tensor computation for better programmability and extensibility.
    \item Extensive experiments demonstrate the advantages of \Mname~in terms of better performance compared with the state-of-the-art DGL framework on mainstream GNN models across various datasets.
\end{itemize}
\section{Background and Related Work}
\label{sect: background and related work}
In this section, we will introduce the background of GNNs, the quantization of GNNs, and basics of GPU Tensor Core. 

\subsection{Graph Neural Networks}
\label{sect: Graph Neural Networks}
\begin{figure} [t] \small
    \centering
    \includegraphics[width=\columnwidth]{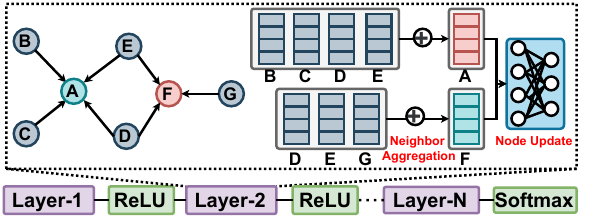}
    \caption{GNN General Computation Flow.}
    \label{fig: GNN Computation Flow}
     \vspace{5pt}
\end{figure}

Graph neural network (GNN) is an effective tool for graph-based machine learning. The detailed computing flow of GNNs is illustrated in Figure~\ref{fig: GNN Computation Flow}. 
GNNs basically compute the node feature vector (embedding) for node $v$ at layer $k+1$ based on the embedding information at layer $k$ ($k \geq 0$), as shown in Equation~\ref{eq: GNN},
\begin{gather} \small \label{eq: GNN}
 \begin{aligned} 
   a_{v}^{(k+1)}  &= \boldsymbol{\mathit{Aggregate}}^{(k+1)}({h_{u}^{(k)}|u\in \mathbf{N}(v)\cup h_v^{(k)}}) \\
   h_{v}^{(k+1)}  &= \boldsymbol{\mathit{Update}}^{(k+1)}(a_{v}^{(k+1)})
\end{aligned}   
\end{gather}
where $h_{v}^{(k)}$ is the embedding vector for node $v$ at layer $k$; $a_{v}^{(k+1)}$ is the aggregation results through collecting neighbors' information (\textit{e.g.}, node embeddings); $\mathbf{N}(v)$ is the neighbor set of node $v$.
The aggregation method and the order of aggregation and update could vary across different GNNs. 
Some methods~\cite{GCNConv,SageConv} just rely on the neighboring nodes while others~\cite{GATConv} also leverage the edge properties that are computed by applying vector dot-product between source and destination node embeddings. The update function is generally composed of standard NN operations, such as a single fully connected layer or a multi-layer perceptron (MLP) in the form of $w\cdot a_{v}^{(k+1)} + b$, where $w$ and $b$ are the weight and bias parameter, respectively. The common choices for node embedding dimensions are 16, 64, and 128, and the embedding dimension may change across different layers.

The most recent advancement of GNN is its batched computation~\cite{chiang2019cluster}, which has also been adopted by many state-of-the-art GNN computing frameworks~\cite{wang2019dgl, pyG} for large graphs that cannot be easily fit into the GPU/CPU memory for computation directly.
Batched GNN computation has been highlighted with good accuracy and runtime performance~\cite{chiang2019cluster} in comparison with full-graph computation.
Batched GNN computation takes several steps. 
\hlp{First}, it decomposed the input graphs by employing the state-of-the-art graph partitioning toolset, such as METIS~\cite{METIS}, which can minimize the graph structural information loss meanwhile maximizing the number of edge connections within each subgraph (\textit{i.e.}, improving the subgraph modularity).
\hlp{Second}, it feeds the small subgraphs into the GNN models for computation, which will generate the node feature vector for each subgraph. 
\hlp{Third}, the generated node embeddings can be used in multiple downstream tasks, such as node/graph classification, link prediction, and community detection~\cite{liben2007link, grover2016node2vec, huang2021broader,pole}.

\subsection{Quantization of GNNs}
Besides the research efforts on full-precision GNNs, recent focus also shifts towards the quantized GNNs. 
For example, Boyuan et al.~\cite{feng2020sgquant} propose the first framework for running quantized GNNs, and several types of quantization schemes can be applied on GNNs (\textit{e.g.}, the quantization based on the GNN layer, node degrees, and the edge weights). And their experimental results also demonstrate the effectiveness of the GNN quantization in terms of memory saving and model accuracy. 
Shyam et al.~\cite{tailor2020degree} introduce an architecturally agnostic and stable method, Degree Quant, to improve performance over existing quantization-aware training baselines commonly used on other architectures (\textit{e.g.}, CNNs). They achieve up to $4.7\times$ speedups on CPU when using \texttt{int8} compared with \texttt{float}.
Compared with the full-precision GNNs, low-bit GNNs bring the benefit of model robustness towards the adversarial attacks and the low computation and memory overheads. 
However, work from~\cite{feng2020sgquant} only showcases the theoretical memory and computation benefits via software-level quantization simulation, where its underlying computation is still carried out in 32-bit full-precision \texttt{float}. 
Work from~\cite{tailor2020degree} only demonstrates such gains on CPUs, which has limited applicability in the real-world GNN computation settings.
This encourages us to harvest its real performance benefits on the modern widely used GPU platforms.

\subsection{Tensor Core on GPUs}
\vspace{1pt}
The recent advancement of GPU hardware technology has pushed computing power to a new level. Among those innovations, the most significant one is the Tensor Core (TC) on NVIDIA GPU. 
Different from scalar-scalar computation on CUDA Cores, TC provides a matrix-matrix compute primitive, which can deliver more than $10\times$ higher computation throughput.
The initial version of TC is designed for handling the GEMM with half-precision input and full-precision output. 
More variants (\textit{e.g.}, \texttt{int8}, \texttt{int4}, and \texttt{int1} inputs with 32-bit unsigned integer (\texttt{uint32}) output) have been introduced since the recent CUDA release (11.0) and newer GPU microarchitectures (\textit{e.g.}, Turing and Ampere). 

\vspace{1pt}
In particular, TC supports the compute primitive of $\mathbf{D} = \mathbf{A}\times \mathbf{B} + \mathbf{C}$, where matrix tile $\mathbf{A}$ and $\mathbf{B}$ are required to be a certain type of precision (\textit{e.g.}, 1-bit), while matrix tile $\mathbf{C}$ and $\mathbf{D}$ use \texttt{uint32}. 
Depending on the input data precision and the version of GPU microarchitecture, the matrix tile size of $\mathbf{A}(M\times K)$, $\mathbf{B}(K\times N)$, and $\mathbf{C}(M\times N)$ may have different choices. 
For example, 1-bit TC computing requires $M=N=8$ and $K=128$.
Different from the CUDA Cores which requires users to define the execution flow of each thread (\textit{i.e.}, work of individual threads). TC requires the collaboration of a warp of threads (32 threads) (\textit{i.e.}, work of individual warps). 
This can be reflected in two ways. 
\hlp{First}, before calling TC for computation, all registers of a warp of threads need to collaboratively store the matrix tile into a new memory hierarchy (called \textit{Fragment}~\cite{TCProgram}), which allows data sharing across registers. This intra-warp sharing provides opportunities for fragment-based memory optimizations. 
\hlp{Second}, during the computation, these loaded matrix fragments will be taken as the TC input to generate the output fragment, which also consists of the registers from each thread in a warp. Data movements among these registers are also managed by a warp of threads collaboratively.

\vspace{1pt}
Prior research efforts have been devoted to accelerating high-performance computing workloads with TC. 
Ahmad et al.~\cite{BatchTCMM} process the batched small-size GEMM on TC for acceleration.
Ang and Simon~\cite{Li2020AcceleratingBN} leverage 1-bit GEMM capability on Turing GPU TC for accelerating binary neural network inference.
Dakkak et al.~\cite{reduction} accelerates the half-precision scan on TC by transforming the scan to a GEMM.
Boyuan et al.~\cite{EMM-TC} introduce GEMM-based scientific computing on TC with extended precision.
\Mname~enlarges the application range of TC by accelerating GNNs for any-bitwidth quantized GNN computation, which is not directly covered by any existing research, any release of cuBLAS~\cite{cublas}, or CUTLASS~\cite{cutlass} library, and GPU TC hardware.
\begin{figure}[!t]
\centering
\begin{lstlisting}[caption=Basic WMMA APIs for TCU in CUDA C.,style=tt1,label=code: wmma API interface.]
// define the register fragment for matrix A (1-bit).
wmma::fragment<matrix_a, M, N, K, b1, row_major> a_frag;
// load a tile of matrix A to register fragment.
wmma::load_matrix_sync(a_frag, A, M);
// matrix-matrix multiplication (1-bit x 1-bit -> 32-bit)
wmma::mma_sync(c_frag, a_frag, b_frag, c_frag);
// store the C matrix tile from register to matrix C.
wmma::store_matrix_sync(C, c_frag, N, mem_row_major);
\end{lstlisting} 
\end{figure}

\vspace{1pt}
TC can be used in several ways. 
The simplest one is to call cuBLAS~\cite{cublas} \texttt{cublasSgemmEX} API.
However, cuBLAS API only supports computation on the most common fixed bit-width on TC, such as 8-bit, half-precision (16-bit), thus, it cannot support any bitwidth precision directly. 
The second way is to call the Warp Matrix Multiply-Accumulate (WMMA) (\texttt{nvcuda::wmma}) API~\cite{wmma} in CUDA C++ to operate TC directly. There are basically four types of operations (Listing~\ref{code: wmma API interface.}). 
In this project, we follow the second way for more low-level implementation customization for batched GNN computation. Because it can offer more design/implementation flexibility for compositing arbitrary-bit computation and ease the optimization (\textit{e.g.}, data loading and reuse) for batched GNN-specific workloads at the GPU kernel.  
\vspace{-10pt}
\section{\Mname~Algorithm Design}
In this section, we first introduce the basics of low-bit computation. Then we will discuss our TC-tailored algorithm design for quantized GNN.

\subsection{1-bit Composition for Quantized Ops.}
Over the last few years, quantized deep neural networks (QDNNs)~\cite{feng2020sgquant,tailor2020degree} have been extensively studied, largely due to their memory saving and high computation performance. 
In GNN, however, similar work is largely lagging behind. Work from \cite{feng2020sgquant}~demonstrates that GNN is actually insensitive to quantization, even very low-bit quantization would not lead to evident accuracy loss because of the graph-like aggregation operations that can amortize such quantization influence. Another work from~\cite{bahri2020binary}~also demonstrates that even the binarized GNN would be beneficial in some application scenarios. 
In this work, we foresee that the support for any-bitwidth precision computation on GNN is vital to satisfy various users' demands (\textit{e.g.}, execution time). 

Given a quantization bit $q$ and the 32-bit floating-point value $\alpha\in \mathcal{R}$, we quantize it as a $q$-bit value by using 
\begin{equation} \small \label{eq:scale_quantize}
    \alpha^{(q)} = \floor*{\frac{\alpha - \alpha_{min}}{scale}}.
\end{equation}
where $\alpha_{min}$ is an empirical lower bound that can be determined by users or application settings; $scale$ is the ratio between the range ($|\alpha_{max} - \alpha_{min}|$, where $\alpha_{max}$ is an empirical upper bound) and the $q$-bit representation range ($2^q$); $\floor*{\cdot}$ is the floor function.

For any-bitwidth computation on quantized values, we propose a new type of arithmetics based on the ``atomic'' 1-bit computation widely used in the binarized NN~\cite{hubara2016binarized}. 

\textbf{Any-bitwidth Scalar-Scalar Multiplication:} Assuming we have a 3-bit scalar value ($a$) and multiply it with a 2-bit scalar value ($b$). we can first represent these two values as 
\begin{equation}
\begin{array}{l}
    a = at_2\cdot 2^{2} + at_1\cdot 2^{1} + at_0 \cdot 2^{0} \\
    b = bt_1\cdot 2^{1} + bt_0 \cdot 2^{0}
\end{array}
\end{equation}
where $at_*$ and $bt_*$ indicate the bit value (0/1) at the certain bit position after bit decomposition.
By following the general rule of multiplication, we can get $a\cdot b$ as
\begin{equation}
 a\cdot b = (at_2\cdot 2^{2} + at_1\cdot 2^{1} + at_0 \cdot 2^{0}) (bt_1\cdot 2^{1} + bt_0 \cdot 2^{0})\\
\end{equation}
through simplification we can get that 
\begin{equation}
\begin{array}{l}
 a\cdot b = at_2bt_1\cdot 2^{3} + (at_1bt_1 + at_2bt_0)\cdot 2^{2} \\ 
  \hspace{20pt} + (at_0bt_1 + at_1bt_0) \cdot 2^{1} + at_0bt_0\cdot 2 ^{0}
\end{array}
\end{equation}

\textbf{Any-bitwidth Vector-Vector Multiplication:} 
We extend the any-bitwidth scalar-scalar computation towards any-bitwidth vector-vector computation between a 3-bit vector $\overrightarrow{v_i}$ and 2-bit vector $\overrightarrow{v_j}$, each of which has $k$ elements. Therefore, the above scalar-scalar multiplication formula can be extended to $k$-dimension vector-vector multiplication
\begin{equation}
    \begin{array}{l}
     \overrightarrow{v_i}\cdot\overrightarrow{v_j} = 
     \sum\limits^{k}_y a^{(y)}\cdot b^{(y)} = 
     \sum\limits^{k}_y
     at_2^{(y)}bt_1^{(y)}\cdot 2^{3} \\ 
      \hspace{20pt} +  \sum\limits^{k}_y (at_1^{(y)}bt_1^{(y)} + at_2^{(y)}bt_0^{(y)})\cdot 2^{2} \\ 
      \hspace{20pt} +  \sum\limits^{k}_y (at_0^{(y)}bt_1^{(y)} + at_1^{(y)}bt_0^{(y)}) \cdot 2^{1} +  \sum\limits^{k}_y at_0^{(y)}bt_0^{(y)}\cdot 2^{0}
    \end{array}
\end{equation}
From the above formula, we can see that in order to compute the result of any-bitwidth vector-vector multiplication, we first do bit decomposition on all elements in each vector then do bit-bit multiplication between elements from each vector, and finally do bit shifting and reduction to get the final result.
For example, after bit-decomposition of $\overrightarrow{v_i}$ and 
$\overrightarrow{v_j}$, we can get $\overrightarrow{v_i}$ at bit position 2 as $at_2^{(y)}$ and $\overrightarrow{v_j}$ at bit position 1 as $bt_1^{(y)}$, where $y\in [0, k)$. 
From the multiplication and addition, we can get the multiplication result of $\overrightarrow{v_i}\cdot\overrightarrow{v_j}$ at bit position 3. Such a 1-bit vector-vector multiplication can be effectively implemented as 
\begin{equation}
    ans_{i,j} = \mathit{popcnt}(\overrightarrow{v_i} \& \overrightarrow{v_j})
\end{equation}
where $popcnt()$ counts the total number of $1s$ of the result in its bit representation (\textit{e.g.}, $popcnt$ will return $3$ for a binary number $1011$).
A similar procedure can be applied to generate the result at bit position 0, 1, and 2. After all these individual bits in temporary results are ready, we can do bit shifting and reduction to get the final result. Based on such any-bitwidth vector-vector results, we can easily derive the any-bit matrix-matrix multiplication scheme, where each element in the output matrix can be seen as the results of any-bitwidth vector-vector multiplication. 

\subsection{Quantized Computation in GNNs}
Applying any-bitwidth precision computation in GNNs would find two major specialties. 
\hlp{First}, the adjacent matrix ($\mathbf{A}$) of GNNs only need to use binary (1-bit) number to represent the presence/absence of edges. 
\hlp{Second}, the node embedding matrix ($\mathbf{X}$) and the weight matrix ($\mathbf{W}$) can be represented with any-bitwidth to meet the different precision demands. 
 \begin{algorithm}[t] \footnotesize 
  \caption{1-layer Quantized GNN Computation.}
  \label{algo: APGNN algorithm}
\SetAlgoLined
  \SetKwInOut{Input}{input}
  \SetKwInOut{Output}{output}
  \Input{Full-bit adjacent matrix $\mathbf{A}$ ($N\times N$), node embedding matrix $\mathbf{X}$ ($N\times D$), and weight matrix $\mathbf{W}$ ($N\times H$).}
  \Output{Updated full-bit node embedding matrix $\mathbf{\hat{X}}$ ($N\times H$).}
    \tcc{Bit decomposition of the input matrices.}
    $\mathbf{A_{bin}}$ = \textbf{bitDecompse}($\mathbf{A}$, 1)[0]\;
    $X\_list$ = \textbf{bitDecompse}($\mathbf{X}$, s)\;
    $W\_list$ = \textbf{bitDecompse}($\mathbf{W}$, t)\;
    $X\_new\_list$ = list(); $C\_dict$ = dict(); $\mathbf{\hat{X}} = zeros\_as(\mathbf{X})$\;
    \tcc{Neighbor aggregation by bit-GEMM ($\mathbf{A}\times\mathbf{X}$).}
    \For{$xIdx$ \textbf{in} \textbf{len}($X\_list$)}{
         $X\_new\_list$.append(\textbf{BMM}($\mathbf{A_{bin}}$, $X\_list$[$xIdx$]))\;
    }
    \tcc{Node update by bit-GEMM ($\mathbf{X\_{new}}\times\mathbf{W}$).}
    \For{xIdx \textbf{in} \textbf{len}($X\_new\_list$)}{
        \For{wIdx \textbf{in} \textbf{len}($W\_new\_list$)}{
            \tcc{Compute bit-matrix at target bit level.}
            $bitIdx = xIdx + wIdx$\;
            $tmp\_C$ = \textbf{BMM}($X\_new\_list$[$xIdx$], $W\_list$[$wIdx$])\;
            $C\_dict$[$bitIdx$].append($tmp\_C$)\;
        }
    }
    \tcc{Elementwise reduction of results.}
    \For{bitIdx \textbf{in} \textbf{len}(C\_dict)}{
        \For{Idx \textbf{in} \textbf{len}(C\_dict[bitIdx])}{
            $\hat{X}$[$Idx$] += $C\_dict$[$bitIdx$][$Idx$] $\ll$ ${bitIdx}$\; 
        }
     }
\end{algorithm}
As described in Algorithm~\ref{algo: APGNN algorithm}, each layer of any-bitwidth GNN consists of {a \textit{neighbor aggregation} and a \textit{node embedding update}} phase. Specifically, neighbor aggregation conducts $\mathbf{X\_new}=\mathbf{A}\cdot\mathbf{X}$ through a \textit{1-bit-and-s-bit} matrix multiplication and the node update conducts $\mathbf{\hat{X}} = \mathbf{X\_new}\cdot\mathbf{W}$ through a \textit{s-bit-and-t-bit} matrix multiplication.
{At Line 1 to 3, we do \textbf{bitDecompose} for subgraph adjacency matrix $\mathbf{A}$, embedding matrix $\mathbf{X}$, and
weight matrix $\mathbf{W}$. For scalar \texttt{int32} numbers, our \textbf{bitDecompose} will first quantize it to another \texttt{int32} number in a n-bit data range $[0, 2^n-1]$ by using Equation~\ref{eq:scale_quantize}. 
Then, it applies bit-shifting to extract each bit (0/1) from the quantized \texttt{int32} number. 
Our 3D stacked bit compression (Section~\ref{sect: 3D-stacked Bit Compression}) happens after the above first and second steps are applied to each element of a matrix, and it will pack the extracted bits for the whole matrix together. 
Here for ease of algorithm description, we maintain different bits of a matrix as the list, \textit{e.g.}, $\mathbf{X}[1]$ stands for the 0's bits for all elements inside the $\mathbf{X}$. At Line 5 to 7, we apply bit-matrix multiplication between each bit matrix from $\mathbf{X}$ and the binary 1-bit matrix $\mathbf{A_{bin}}$, the results of this step will still be a set of bit matrices and be stored in a list. At Line 8 to 14, we apply the similar bit-matrix multiplication between $\mathbf{X}$ and $\mathbf{W}$, and the results of this step will be stored as bit-matrix as well for the following final-result generation.}
To avoid any data overflow during the reduction (Line 15 to 19), $\mathbf{\hat{X}}$ should also use a full-bit data type (\textit{e.g.}, \texttt{int32}).
For large graphs, their adjacent matrices cannot be easily fit into the GPU device memory directly. In this scenario, we employ METIS~\cite{METIS} for graph partitioning and run GNN as batched subgraph computation, which is used by the most popular cluster-GCN~\cite{chiang2019cluster} design. 
Considering that the number of subgraphs generated by METIS~\cite{METIS} is usually within the reasonable size (2,000 to 20,000), such a batched GNN computation can be accommodated on a single modern GPU without {{violating its memory constraints.}}  
Note that to reduce the runtime overhead, the bit-decomposition of the matrix $\mathbf{W}$ and $\mathbf{A}$ can be pre-computed and cached before the GNN computation at each layer. The major reason behind this is that across different GNN layers of the same subgraphs, the adjacent matrix $\mathbf{A}$ can be reused. On the other side, across different subgraphs at the same GNN layers, the weight matrix $\mathbf{W}$ can be reused for the later-on computation.
\section{Implementation}
\subsection{Subgraph Partitioning and Batching}
Real-world graphs usually come with a large number of nodes and highly irregular graph structure (edge connections). 
This brings two levels of difficulties for GNN computing. The first one is the memory consumption, since GPU device memory cannot accommodate all nodes, edges, and node embedding features at the same time. The second one is the inefficient execution since the irregular and sparse edge connections lead to low memory access efficiency and poor computation performance.
To this end, in \Mname, we combine the state-of-the-art graph partitioning technique METIS~\cite{METIS} and subgraph batch processing~\cite{chiang2019cluster} to handle different sizes of input graphs effectively.
Compared with other solutions, such as graph clustering approaches~\cite{raghavan2007near, karantasis2014parallelization} and BFS-based methods~\cite{RCM-Algorithm}, METIS achieves a better quality of its captured subgraph partitions (more edges in each subgraph) and the significantly higher runtime performance owing to its partial parallelization. Note that the number of subgraphs/partitions is determined by users and is passed as a runtime parameter to METIS.

After the subgraph partitioning, we will conduct a batching step for GNN computation on GPUs. This step gathers a set of subgraph partitions based on user-defined batch size. 
Later, during the GNN computing, subgraphs are loaded to GPU memory by batch. 
Using the partitioning and batching strategy for GNN computing gives users control of workloads at two levels of granularity. 
\hlp{First}, the workload granularity is defined by the number of subgraphs/partitions. This would manage the size of each subgraph partition and the edge connection density of each subgraph. In general, the more number of the subgraphs/partitions would lead to denser edges connections within each subgraph, which may bring better computation and memory locality.
\hlp{Second}, the processing granularity is controlled by the batch size. 
This would determine the size of graphs that will be fit into the GPU at each round of execution. The selection of batch size would maximize the utilization of the GPU while respecting the GPU computation and memory resource constraints.

\vspace{-3pt}
\subsection{3D-Stacked Bit Compression}
\label{sect: 3D-stacked Bit Compression}
Existing NN frameworks are developed for full-precision computation, which leads to two major challenges: 
\hlp{First}, the low-bit quantized data type cannot directly borrow the full-precision data type as the ``vehicle'' for computation. The major reason is that the full precision data type such as $\texttt{float}$ and $\texttt{int32}$ cannot bring any benefits to the memory or computation saving. 
\hlp{Second}, low-bit quantization would not fit any type of bit alignment, since its bit-level boundary mostly cannot be divisible by the size of a byte (8-bit), making it hard to retrieve its actual value. 

\begin{figure}[t] \small
    \centering
    \includegraphics[width=\linewidth]{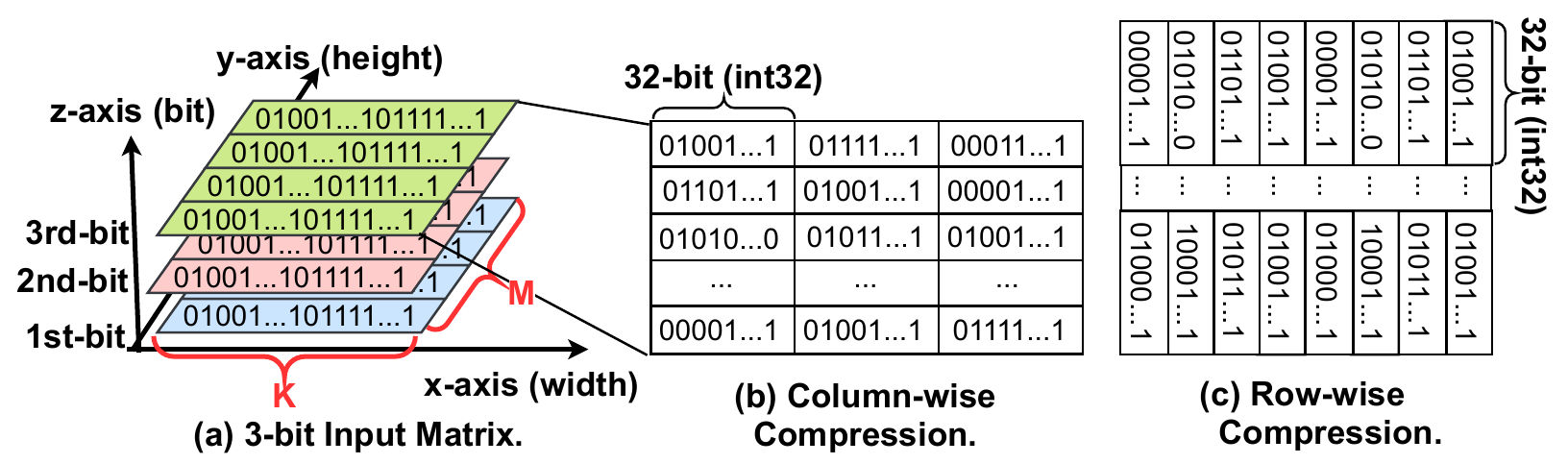}
    \caption{3D-Stacked Bit Compression. Note that every 32 bits are compressed and stored in little-endian.}
    \label{fig: bit layout}
    \vspace{6.6pt}
\end{figure}

To this end, we propose a novel \textit{3D-stacked bit-compression} technique to handle any-bitwidth data type effectively. The major idea is to compress any-bitwidth input with 32-bit alignment for ease of value retrieval and memory alignment. As exemplified in Figure~\ref{fig: bit layout}(a), we have an input matrix with the shape of 3-bit$\times M\times K$. For each bit of the element in the matrix, we store it in a bit matrix (1-bit$\times M\times K$) stacked along the vertical $z$ axis. During the computation of any-bitwidth matrix multiplication $\mathbf{C} = \mathbf{A}\times\mathbf{B}$, two types of 3D-stacked bit-compression are employed. For matrix $\mathbf{A}$, we use the \textit{column-wise compression} with 32-bit alignment, as illustrated in Figure~\ref{fig: bit layout}(b).
The main reason for choosing column-wise compression is that the matrix multiplication would benefit from coalesced across-column memory access along each row of the matrix $\mathbf{A}$. 32-bit alignment can benefit the read performance by coalesced loading from the global memory to fragment. After the compression on matrix $\mathbf{A}$ (1-bit$\times M\times K$), we will get a 32-bit compressed 3-bit $\mathbf{A_c}$ with the shape of 3-bit$\times(\mathbf{PAD8}(M)\times\floor{\mathbf{PAD128}(K)/32})$, where $\mathbf{PAD8}$ and $\mathbf{PAD128}$ are for padding rows/columns that cannot be divisible by the basic TC computing size ($M(8)\times N(8)\times K(128)$).
For matrix $\mathbf{B}$, we use the \textit{row-wise compression} with 32-bit alignment, as shown in Figure~\ref{fig: bit layout}(c) which can benefit the across-row access along each column of matrix $\mathbf{B}$. After the compression on matrix $\mathbf{B}$ (1-bit$\times M\times$K), we will get a 32-bit compressed 2-bit $\mathbf{B_c}$ with the shape of 2-bit$\times\floor{\mathbf{PAD128}(K)/32}\times\mathbf{PAD8}(N)$ for the output layer. Note that if the $\mathbf{A}\times \mathbf{B}$ is the hidden layer of a GNN model, the padding strategy on matrix $\mathbf{B}$ would be slightly different considering that the result matrix $\mathbf{C}$ will become a new matrix $\mathbf{A}$ in the next layer which demands 128-bit padding. In this case, to avoid additional padding overhead, we will get the 2-bit $\mathbf{B_c}$ with the shape of 2-bit$\times\floor{\mathbf{PAD128}(K)/32}\times\mathbf{PAD128}(N)$. 

Compared with the previous work~\cite{cowan2020automatic} that also leverages bit-level data packing, there are several differences. 
The \hlp{first} one is the \textit{padding strategy}. Padding of \Mname~on different tensor dimensions could be different, where bit-level padding is ignored in the work from \cite{cowan2020automatic}. 
For example, QGTC may PAD8 or PAD128, depending on the following computation is carried out in low-bit or 32-bit format, thereby, avoiding unnecessary conversion. 
The \hlp{second} one is the \textit{packing datatype}. Work from \cite{cowan2020automatic}~uses \texttt{uint4} for packing continuous 128 bits, while QGTC uses a 32-bit format for better interoperability with PyTorch.
The \hlp{third} one is the \textit{bit-level layout}. Work from \cite{cowan2020automatic} doesn’t consider more bit-level layout optimization. In QGTC, for GEMM operation ($\mathbf{C=AX}$), we use a column-wise compression for the matrix $\mathbf{A}$ and a row-wise compression for the matrix $\mathbf{X}$.

\vspace{-5pt}
\subsection{Zero-tile Jumping}
Even though the subgraph partitioning such as METIS~\cite{METIS} makes the subgraph denser (more number of edge connections within each subgraph), there are still some TC tiles (\textit{i.e.}, the input matrix tile for a single TC computation) that are filled with all-zero elements. 
Therefore, directly iterating through these zero tiles would introduce the cost of unnecessary memory (loading data from the global memory to thread-local registers) and computation (processing 1-bit TC-GEMM on input adjacent matrix tile that contains all-zero elements). 
Based on this observation, we introduce a novel zero-tile jumping technique to reduce the unnecessary computations by bitwise \texttt{OR} operations and warp-level synchronization primitives. 
\begin{figure}[t] \small
    \centering
    \includegraphics[width=\columnwidth, height=4cm]{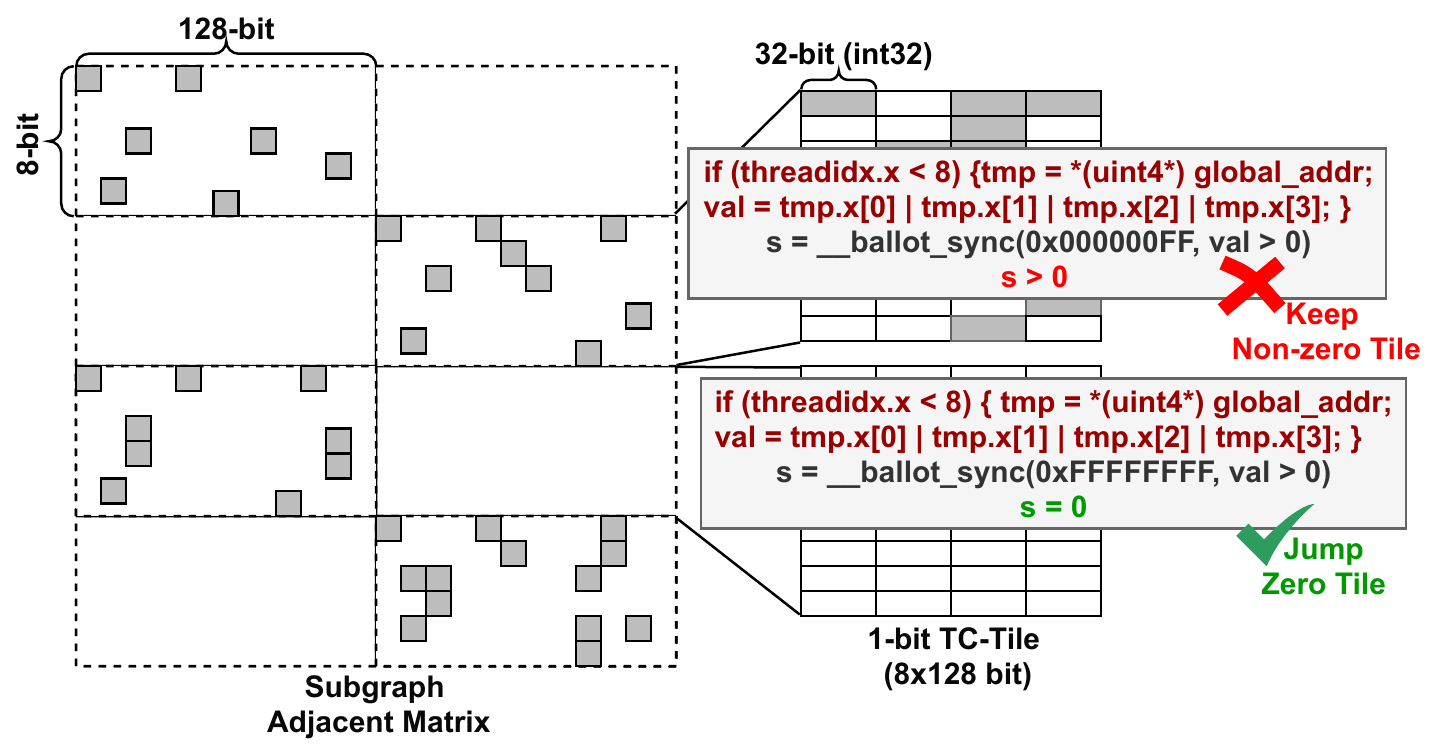}
    \vspace{-5pt}
    \caption{Zero-tile Jumping. Note that each small grey square box (on the left side) indicates an edge connection between two nodes within a graph. Each grey rectangular box (on the right side) indicates at least one of its 32 consecutive small square boxes is grey (the presence of an edge).}
    \label{fig: Zero Tile Jumping.}
    \vspace{5pt}
\end{figure}

As illustrated in Figure~\ref{fig: Zero Tile Jumping.}, each 1-bit TC-GEMM would work on the tile size of $8\times128$ register fragment. 
This can be well partitioned into $8\times4$ \texttt{int32} elements. 
To check whether the $8\times128$ tile contains all-zero elements, we first employ only 8 threads from a warp of threads to fetch an \texttt{uint4\_v} vector data (each \texttt{uint4\_v} element in CUDA consists of 4 \texttt{int32} elements placed in continuous memory address). 
The reason for using \texttt{uint4\_v} is to maximize the memory access efficiency by issuing fewer global memory requests.
Once all \texttt{uint4\_v} elements have been loaded. 
Each thread will apply bitwise \texttt{OR} across all 4 \texttt{int32} elements, which will check whether each row of a TC-tile is all-zero. 
The next step is to tell whether the whole tile is all-zeros across different rows, we will use the warp-level primitive to sync the information across these 8 active threads in the warp. This step will generate an \texttt{int32} number. If this number is zero, it will indicate all elements in this input TC-tile are zero. Otherwise, we still need to conduct the 1-bit TC-GEMM on the current tile.  
We will give a more quantitative analysis of such zero-tile jumping in Section~\ref{sect: Additional Studies}.

\vspace{-5pt}
\subsection{Non-Zero Tile Reuse}
\begin{figure}[t] \small
    \centering
    \includegraphics[width=\columnwidth, height=2.5cm]{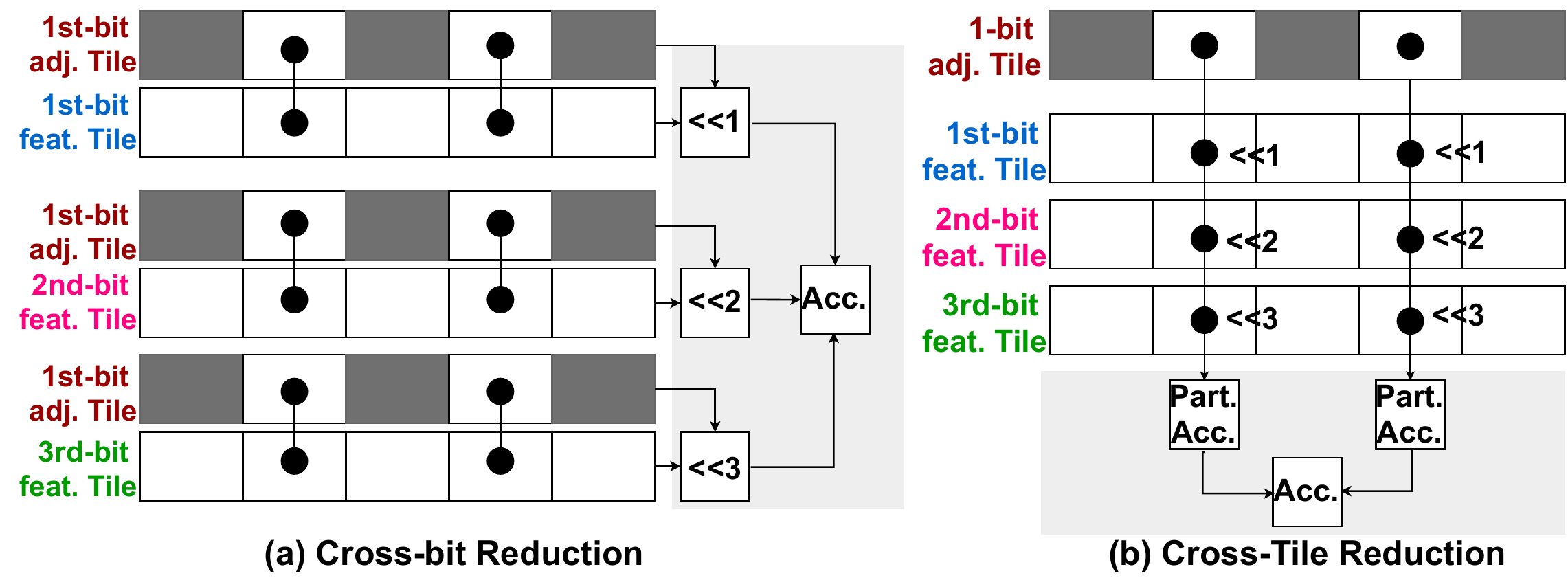}
    \vspace{-10pt}
    \caption{Non-zero Tile Reuse. Note that the grey box indicates the zero-tile of the subgraph adjacent matrix, while the white box with a block solid dot inside represents the non-zero tiles of the subgraph adjacent matrix.}
    \label{fig: Non-zero Tile Reuse.}
    \vspace{10pt}
\end{figure}

In addition to jumping over the zero tiles, we further consider reusing the non-zero tiles to improve data locality.
In the aggregation step of the GNN computation, we generate the output feature map at different bit-level separately. 
For example, when we choose 1-bit adjacent subgraph matrix and a 4-bit feature embedding matrix, we will execute the iteration 4 times to generate the output. 
One straightforward solution, called \textit{cross-bit reduction}, is to generate the complete output matrix tile at each bit level first. This requires loading the matrix tile imperatively, as shown in Figure~\ref{fig: Non-zero Tile Reuse.}(a). 
However, this would cause one problem that each non-zero tile from the adjacent matrix will be repetitively loaded when computing with each bit matrix from the embedding matrix. 

In fact, we can consider reordering the steps in a way that we can maximize the benefit of each non-zero tile of the subgraph adjacency matrix.
As shown in Figure~\ref{fig: Non-zero Tile Reuse.}(b), we introduce a \textit{cross-tile reduction} strategy. Specifically, for each loaded non-zero fragment, we will use it to generate an output tile at all bit levels and do a localized reduction (only on the current tile) to generate a partial aggregation result. 
Once this part has been done, we will move forward to the next non-zero tile and repeat the same process until all non-zero tiles have been processed. 
%
The complexity of loading the nonzero tiles can be reduced from $\mathcal{O}(n)$ to $\mathcal{O}(1)$, where $n$ is the number of bits for node embeddings.

\subsection{Inter-layer Kernel Fusion}
Across the GNN layers, we incorporate the low-bit data transferring. Specifically, the output of the one hidden layer will directly be handed over to the next layer as the input. 
There are several strategies we use. 
\hlp{First}, we apply data quantization and bit-decomposition at the end of the computation kernel such as the neighbor aggregation and node update. This can help to avoid outputting the result to the slow global memory and apply the data manipulation again. 
\hlp{Second}, standalone activation function kernels such as \texttt{ReLU} and \texttt{tanh}, can be effectively fused into our computation kernel as a \texttt{device} function, which can directly operate the shared memory results to achieve high performance. 
For the batch normalization (BN) layers that follow the graph convolution layers, we can also do layer fusion based on
\begin{equation}
    \mathbf{BN}(x_{i,j}) = \left( \frac{x_{i,j} - \mathbb{E}[x_{*,j}]}{\sqrt{Var[x_{*,j}] + \epsilon}} \right) \cdot \gamma_j + \beta_j
\end{equation}
where $\beta_j$, $\gamma_j$, and $\epsilon$ are the BN parameters that can be incorporated into the low-bit convolutional kernel to avoid launching a BN kernel. 
One thing worth noting is that computation at the hidden layer and the output layer is slightly different. For hidden layers, each kernel has the quantization + bit-decomposition on the output activation, since the next layer relies on the low-bit data as the input for computation. While for the last layer, once the full-precision accumulation is complete, it will directly output the full-precision result for the \texttt{softmax} layer (considering the node classification task) to generate logits that demand high precision.

\subsection{Bandwidth-Optimized Subgraph Packing} 
During the GNN computation of the subgraphs, data communication between the CPU host and GPU device is also unavoidable. It will swap the subgraph data (such as edge lists and node embedding) in/out of the GPU device. 
One basic approach is to transfer the dense adjacent matrix in floating point numbers considering that the size of a single subgraph is generally within the range of the modern GPU memory. 
%
However, this could easily lead to a huge amount of data traffic between the CPU and GPU host. The transferring performance is heavily bounded by PCIe bandwidth (32 GB/s for PCIe 4.0x16) between the CPU host and the GPU device. For the node embedding matrix, the current practice is to transfer the node embedding matrix by initializing another standalone PCIe transferring, which incurs additional overheads and is unable to maximize the bandwidth usage.

To overcome these issues, we employ a new strategy, called \textit{bandwidth-optimized subgraph packing}. 
Instead of directly migrating the dense adjacent matrix or sparse adjacent matrix in single-precision floating point, we just transfer the compressed low-bit adjacent matrix and low-bit embedding matrix. 
This can significantly minimize the data traffic on the high-cost PCIe-based data communication. Besides, we compress the low-bit adjacent matrix and low-bit embedding matrix into a compound memory object (by using \texttt{torch.nn.Module} and \texttt{register\_buffer}) on the host first and then initiate the transferring of this memory object from the host CPU to GPU device.
\section{Integration with PyTorch}
Besides the highly efficient kernel design and data transferring optimization,
for better usability and programmability, we integrate \Mname~with the popular PyTorch framework. However, there are two key technical challenges. 
The first one is how to represent the quantized low-bit number in those Tensor-based frameworks that are built on byte-based data types (\textit{e.g.}, \texttt{int32}). 
The second one is how to apply valid computation between the quantized low-bit number and those well-defined byte-based numbers. For example, how could we get the correct results when we do arithmetic multiplication between a 2-bit number and a 32-bit integer number. To this end, we introduce two new techniques.

\textbf{Bit-Tensor Data Type:}  
We use the 32-bit {\texttt{IntTensor}} in PyTorch as the ``vehicle'' for holding any-bitwidth quantized data. And we leverage our 3D-stacked bit compression technique (Section~\ref{sect: 3D-stacked Bit Compression}) to package the quantized data. 
We offer a PyTorch API {\texttt{Tensor.to\_bit(nbits)}} for such data type conversion functionality.
Note that existing PyTorch APIs, such as {\texttt{print}}, are only defined for those complete data types, such as Int. Therefore, to access the element value of a bit-Tensor, we provide {\texttt{Tensor.to\_val(nbits)}} to decode a bit-Tensor as \texttt{{int32}} Tensor (converting each element from a low-bit number to an \texttt{int32} number). This can make our design compatible with existing PyTorch functionalities.

\vspace{2pt}
\textbf{Bit-Tensor Computation:} 
We handle two different types of computation: 
1) the operations that only involve bit-Tensor and 2) the operations that involve both bit-Tensor and \texttt{float} or \texttt{int32} Tensor. 
For the first type of operations, we built two APIs based on whether we want to get the \texttt{int32} output or still get the quantized low-bit output as a bit Tensor. 
For any-bitwidth MM with low-bit output, the API is \texttt{bitMM2Bit(C, A, B, bit\_A, bit\_B, Bit\_C)}, where \texttt{A} and \texttt{B} are bit Tensors, \texttt{bit\_A/B/C} are bitwidth parameters. 
For any-bitwidth MM with \texttt{int32} output, the API is
\texttt{bitMM2Int(C, A, B, bit\_A, bit\_B)}.
For the second type of operations, we will first decode a bit-Tensor as a \texttt{float}/\texttt{int32} Tensor by using \texttt{Tensor.to\_val(nbits)}. Then we call the official APIs in PyTorch for the regular full-precision computation.
\begin{table}[t] \small	
\caption{Datasets for evaluation.}
\vspace{-10pt}
\centering
\scalebox{0.93}{
 \begin{tabular}{|| c | l r r r r ||}
 \hline
\textbf{Type} & \textbf{Dataset} & \textbf{\#Vertex} & \textbf{\#Edge} & \textbf{Dim.} & \textbf{{\#Class}}\\
\hline
\multirow{2}{*}{\textbf{I}} 
& Proteins	   & 43,471	& 162,088	& 29   & 2 \\
& artist	        & 50,515	& 1,638,396	    & 100 & 12 \\
\hline
\hline
\multirow{2}{*}{\textbf{II}} 
& BlogCatalog	& 88,784	& 2,093,195	    & 128 & 39 \\
& PPI	       & 56,944	    & 818,716	    & 50    & 121    \\
\hline
\hline

\multirow{2}{*}{\textbf{III}} 
& ogbn-arxiv	    & 169,343	    & 1,166,243		    & 128  & 40 \\
& ogbn-products	    & 2,449,029	    & 61,859,140	    & 100  & 47 \\
\hline
\end{tabular}}
\label{table: Evaluation Dataset}
\vspace{5pt}
\end{table}
\section{Evaluation}
\vspace{2pt}
\begin{figure*}[t] \small
    \centering
    \subfloat[]{\includegraphics[width=0.33\textwidth, height=2cm, trim=0 0cm 0 0 ]{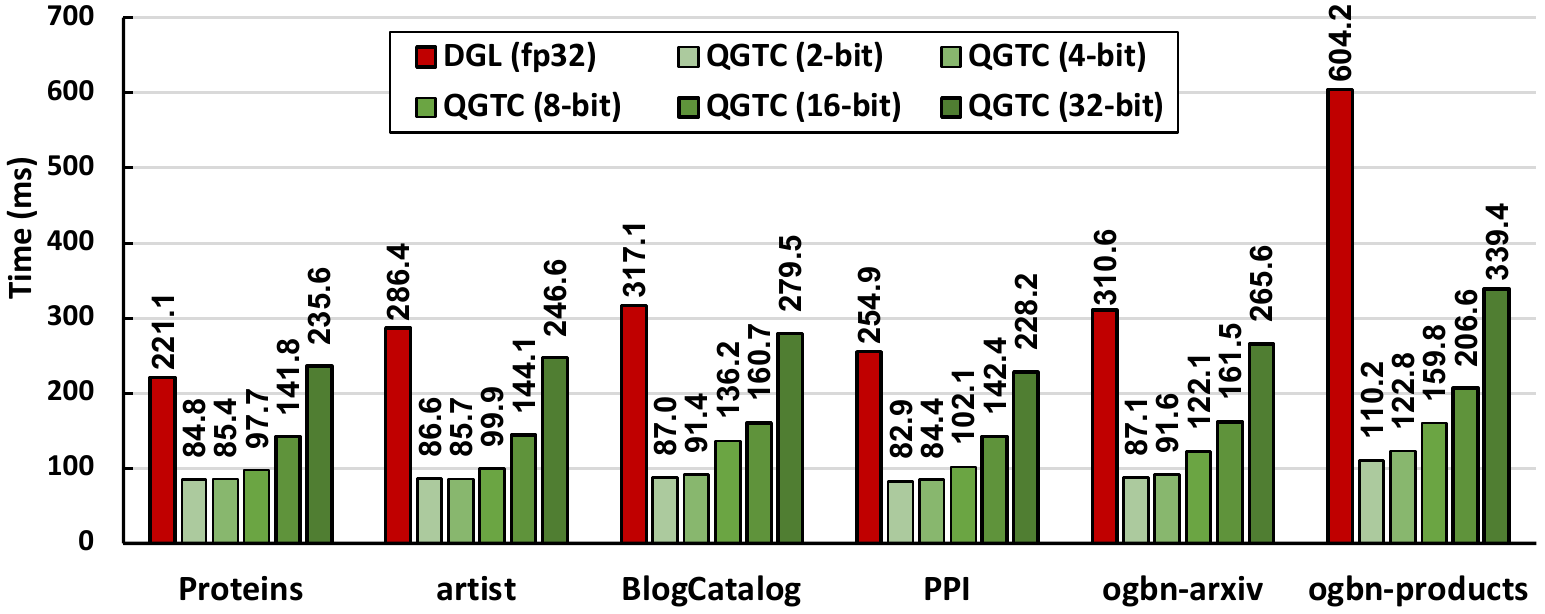}}
    \subfloat[]{\includegraphics[width=0.33\textwidth, height=2cm, trim=0 0cm 0 0cm]{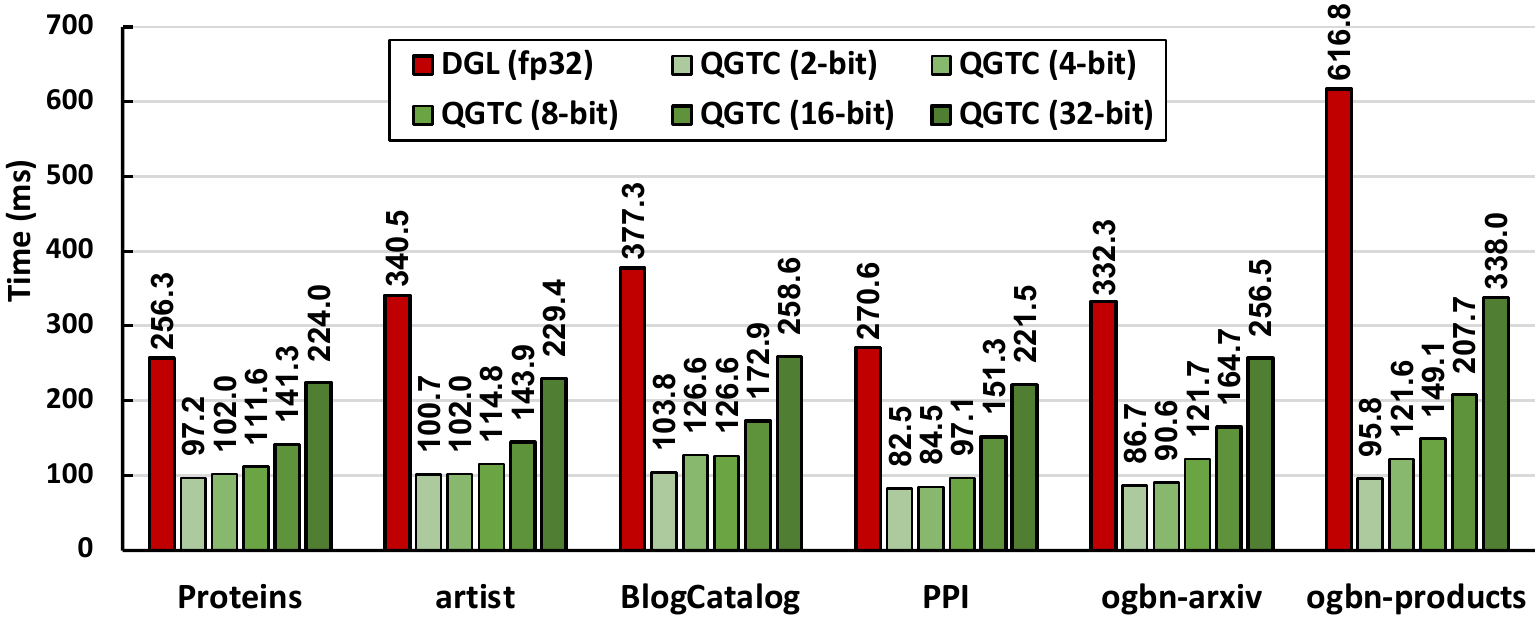}}
    \subfloat[]{\includegraphics[width=0.33\textwidth, height=2cm, trim=0 0cm 0 0cm]{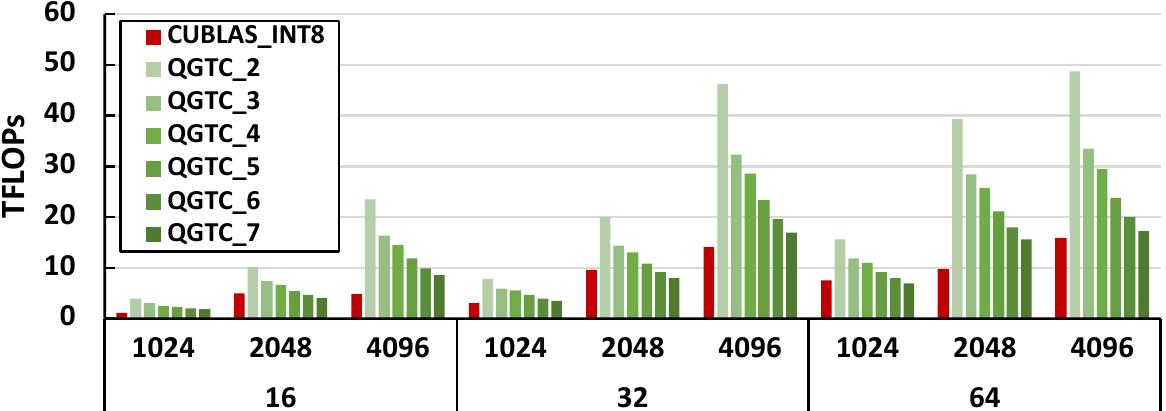}}
    \vspace{-8pt}
    \caption{End-to-end performance comparison with (a) DGL on Cluster GCN and (b) DGL on Batched GIN. (c) Compared with TC-based \texttt{cuBLASgemmEX} (\texttt{int8}) on GNN aggregation kernel throughput performance (in TFLOPs). Note that ``QGTC\_3'' stands for \Mname~with 3-bit data representation for node embedding matrix.}
    \label{fig: overall comparison}
    \vspace{-5pt}
\end{figure*} 

\vspace{2pt}
\textbf{Benchmarks: }
We choose two most representative GNN models widely used by previous work~\cite{wang2019dgl,pyG,ma2019neugraph} on the node classification task to cover different types of aggregation.
{\textbf{1) Cluster GCN}}~\cite{GCNConv} 
is one of the most popular GNN model architectures. 
It is also the key backbone network for many other GNNs, such as GraphSAGE~\cite{SageConv}, and differentiable pooling (Diffpool)~\cite{diffpool}. 
For Cluster GCN evaluation, we use the setting: \textit{3 layers with 16 hidden dimensions per layer.} 
{\textbf{2) Batched GIN}}~\cite{GINConv} differs from cluster GCN in its order of aggregation and node update. 
Batched GIN aggregates neighbor embedding before the node feature update (via linear transformation). GIN demonstrates its strength by capturing the graph properties that cannot be collected by GCN according to~\cite{GINConv}. 
Therefore, improving the performance of GIN will benefit more advanced GNNs, such as GAT~\cite{GATConv}.
For batched GIN evaluation, we use the setting: \textit{3 layers with 64 hidden dimensions per layer}. 
{For quantization bitwidth, we cover the bitwidth settings from the existing quantized GNN studies~\cite{feng2020sgquant,tailor2020degree} and also conduct a comprehensive experimental analysis on different bitwidth settings.}

\textbf{Baselines: } 
In our experiments, we choose several baselines for comparison. 
For end-to-end runtime performance comparison, we choose 
\textbf{Deep Graph Library (DGL)}~\cite{wang2019dgl}, which is the state-of-the-art GNN framework on GPUs. DGL is built with highly optimized CUDA-based GNN kernel as the backend and uses PyTorch~\cite{PyTorch} as its front-end. 
For GNN aggregation kernel performance comparison, we choose the state-of-the-art quantized GEMM implementation on GPU Tensor Core from \textbf{cuBLAS}~\cite{cublas} with \texttt{int8} precision and \textbf{CUTLASS}~\cite{cutlass} with \texttt{int4} precision.

\textbf{Datasets: }
We cover all three types of datasets, which have been used in many previous GNN-related work~\cite{wang2019dgl, pyG}. Details of these datasets are listed in Table~\ref{table: Evaluation Dataset}. Specifically, 
{\textbf{Type I}} graphs are the popular GNN datasets evaluated by many GNN algorithmic papers~\cite{GCNConv, GINConv}. 
{\textbf{Type II}} graphs~\cite{KKMMN2016} are the popular benchmark datasets for graph kernels in many frameworks for GNN algorithmic research. 
{\textbf{Type III}} graphs~\cite{hu2020ogb} are challenging GNN datasets in terms of the large number of nodes and edges. These graphs demonstrate high irregularity in its structures.
Note that we do graph partitioning by using METIS~\cite{METIS} and set the number of total subgraphs as 1,500 as prior work~\cite{chiang2019cluster,zeng2020graphact}.

\textbf{Platforms \& Metrics: } 
\label{sect: Platforms and Metrics }
\Mname~backend is implemented with C++ and CUDA C, while \Mname~front-end is implemented in Python. Our major evaluation platform is a Ubuntu server (16.04) with an 8-core 16-thread Intel Xeon Silver 4110 CPU@2.8GHz with 64GB host memory and an NVIDIA Ampere RTX3090 GPU with 24GB device memory. 
The GPU device kernel is compiled with CUDA (v11.0) and the CPU host code is compiled with GCC 7.5.0 with the compilation option of ``-std=c++14 -O3'' for integration with the PyTorch framework.
To measure the performance speedup, we calculate the averaged latency of 200 rounds of end-to-end results.

\vspace{-10pt}
\subsection{Compared with DGL}
\label{sect: compared with DGL}
In this section, we conduct detailed end-to-end comparison with DGL framework under the different choices of bitwidth. 
As shown in Figure~\ref{fig: overall comparison}(a) and Figure~\ref{fig: overall comparison}(b), \Mname~achieves on average 2.6$\times$ and 2.8$\times$ end-to-end inference speedup compared to DGL over three types of datasets for cluster GCN and batched GIN, respectively. 
We also notice that the performance benefit is closely related to the bitwidth we choose, as we can see that from 16-bit to 32-bit the performance shows a large difference compared with the 1-bit to 8-bit setting.
We next provide a detailed analysis and give insights for each type of dataset.
With a fewer number of bits for both the weights and the node embedding features, \Mname~is more likely to reach higher performance. This is because a smaller size of bitwidth would lead to less memory access and fewer computations at the bit level. 
%
DGL reaches an inferior performance due to 
1) FP32 computation comes with the high computation complexity compared with our \Mname~low-bit design; 
2) DGL can only rely on CUDA cores for computation which is naturally bounded by the peak computation performance compared with our \Mname~on TC with higher throughput performance.
\begin{table}[t]\small
\centering
\caption{Model accuracy \textit{w.r.t.} quantization bitwidth.}
\vspace{-8pt}
\begin{tabular}{|l|r|r|r|r|r|}
\hline
\textbf{Settings} & \textbf{FP32} & \textbf{16 bits} & \textbf{8 bits} & \textbf{4 bits} & \textbf{2 bits} \\ 
\hline
\hline
\textbf{ogb-product} & 0.791 & 0.791 & 0.783 & 0.739 & 0.620 \\ \hline
\textbf{ogb-arxiv} & 0.724 & 0.708 & 0.707 & 0.685 & 0.498 \\ \hline
\end{tabular}
\label{tbl: model accuracy vs. quantization bits}
\vspace{1pt}
\end{table}
Compared with cluster GCN, experimental results on the batched GIN shows higher benefits of \Mname~over DGL. This is because batched GIN applies the node update first before the neighbor aggregation, which leads to higher computation-to-communication ratio.
\Mname~achieves relatively higher performance improvements on Type III datasets. The major reason is that under the same number of partitions, the size of each partition (subgraph) will increase due to more number of nodes/edges.
This also improves the computation intensity that will highlight \Mname's performance advantages of quantized low-bit computation on GPU Tensor cores.

\vspace{2pt}
\textbf{Accuracy \textit{w.r.t.} Quantization Bits} 
To build the QGNN model, we apply quantization-aware training and evaluate the model testing accuracy \textit{w.r.t.} quantization bits on two large Type III datasets on GCN model for demonstration.
As shown in Table~\ref{tbl: model accuracy vs. quantization bits}, the GNN model is resilient to the low-bit quantization and can maintain the model accuracy to a large extent. Combining these results with our above performance evaluation result under different quantization bits, we can conclude that making the right tradeoff between the runtime performance and model accuracy is meaningful and can bring benefits to different application settings.

\subsection{Compared with other baselines}
\textbf{Compared with cuBLAS-\texttt{int8} on TC. }
We further compare our low-bit computation (from 2-bit to 7-bit) with respect to the state-of-the-art \texttt{cuBLASgemmEX} for quantized (\texttt{int8}) GEMM solution on Tensor Core in terms of their throughput performance. Note that \texttt{int8} is the cuBLAS currently supported minimum bits for quantized computation on Tensor Core. In this study, we mainly focus on the computation of $\mathbf{AX}$ (\textit{i.e.}, $N\times N\times D$, where $N$ is the number of nodes and $D$ is the node embedding dimension) for the neighbor aggregation phase. As shown in Figure~\ref{fig: overall comparison}(c), \Mname~achieves significant throughput improvement compared with Tensor Core cuBLAS (\texttt{int8}) in low-bit settings. The major reason is our \Mname~design effectively reduces the computation and the data movements at the bit level, thereby, harvesting the real performance gains of the low-bit quantization on GPUs. When the number of bits for quantization is approaching 8-bit in the computation, the performance gains would decrease due to the increase of bit-level computations.

\vspace{2pt}
\textbf{Compared with CUTLASS-\texttt{int4} on TC}
We also compare against the latest CUTLASS~\cite{cutlass}(v2.7) with the \texttt{int4} Tensor Core GEMM in terms of throughput (TFLOPs) for $\mathbf{AX}$. The results are summarized at Table~\ref{tbl: cmp with cutlass}, where we can clearly see the performance advantage in terms of throughput over the CUTLASS implementation. Note that all reported decimal numbers are in TFLOPS; N is the adjacent matrix size and Dim is the node feature embedding dimension. The graph adjacent matrix is stored in 1-bit. QGTC (2-bit) means the 2-bit representation for the embedding matrix.
\begin{table}[t] \small
\centering
\caption{Compared with CUTLASS-\texttt{int4} (TFLOPs).}
\vspace{-8pt}
\scalebox{0.9}{
\begin{tabular}{|l|l|>{\centering\arraybackslash}p{1.3cm}|>{\centering\arraybackslash}p{1cm}|>{\centering\arraybackslash}p{1cm}|>{\centering\arraybackslash}p{1cm}|>{\centering\arraybackslash}p{1cm}|}
\hline
{\textbf{N}}
& {\textbf{Dim}} 
& {\textbf{CUTLASS (\texttt{int4})}} 
& {\textbf{QGTC (1-bit)}} 
& {\textbf{QGTC (2-bit)}} 
& {\textbf{QGTC (3-bit)}} 
& {\textbf{QGTC (4-bit)}} \\ \hline\hline
2048 & 32 & 10.36 & 32.65 & 19.99 & 14.40 & 11.30 \\ \hline
4096 & 32 & 12.28 & 81.41 & 46.23 & 32.27 & 24.75 \\ \hline
8192 & 32 & 12.67 & 94.58 & 50.82 & 35.22 & 26.31 \\ \hline
2048 & 64 & 21.40 & 63.94 & 39.41 & 29.83 & 22.15 \\ \hline
4096 & 64 & 24.66 & 89.18 & 51.21 & 35.17 & 25.38 \\ \hline
8192 & 64 & 24.70 & 104.66 & 55.16 & 40.77 & 31.07 \\ \hline
\end{tabular}}
\label{tbl: cmp with cutlass}
\end{table}
The major reason behind such performance improvement is that our QGTC design can use the 1-bit binary for representing graph adjacency matrix and n-bit (n=1,2,3,4) for node embedding matrix, while CUTLASS \texttt{int4} only have the support of 4-bit $\times$ 4-bit. 
Thus, we have to use a 4-bit presentation for both adjacent matrix and embedding matrix during computation. 

\subsection{Additional Studies} 
\label{sect: Additional Studies}
In this section, we will conduct detailed studies to demonstrate the effectiveness of our design and optimizations. 

\textbf{Zero-Tile Jumping.} 
We would compute the ratio of the non-zero TC tiles (8$\times$128) that are actually involved in our computation versus the total number of TC tiles in the adjacent matrix.
\begin{figure}[t] \small
    \centering
    \includegraphics[width=0.9\columnwidth]{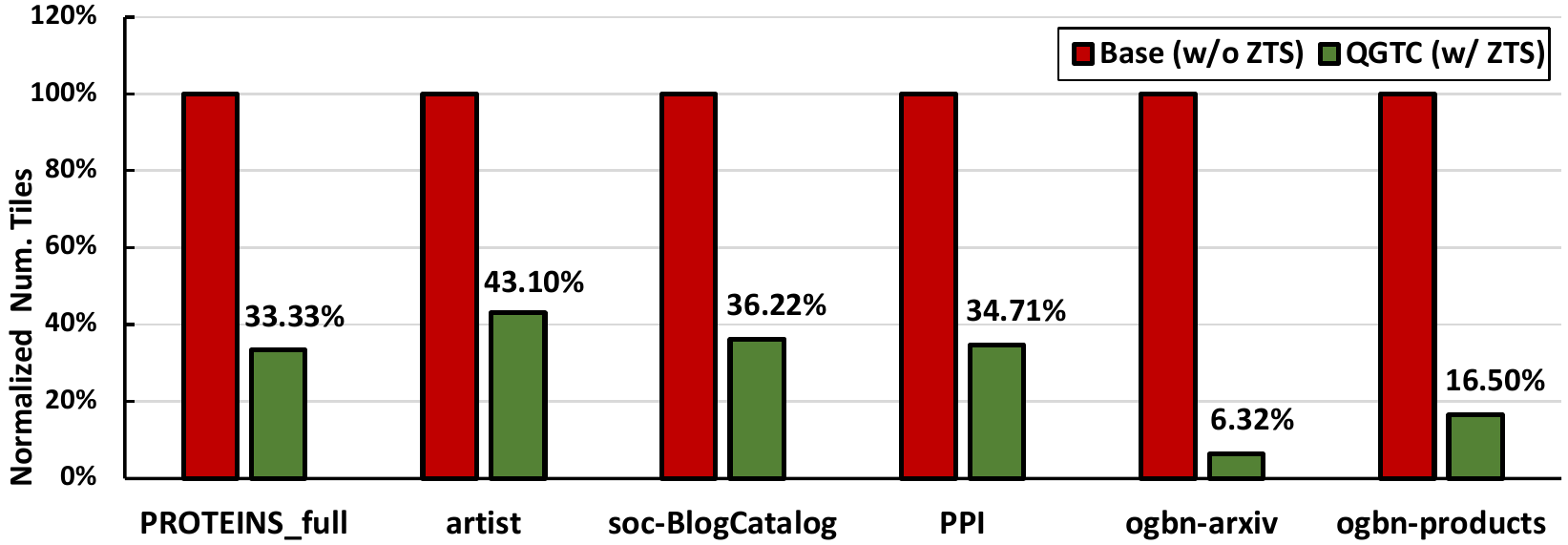}
    \vspace{-3.5pt}
    \caption{Zero-tile jumping efficiency. The percentage (\%) on each green bar indicates the ratio of the number of tiles processed w/ jumping versus w/o jumping solution.}
    \label{fig: Zero-tile jumping efficiency.}
    \vspace{5pt}
\end{figure} 
As shown in Figure~\ref{fig: Zero-tile jumping efficiency.}, our zero-tile jumping technology can largely save the efforts for processing all-zero tiles. 
Based on our observation, the source of such all-zero TC tiles comes from two levels. 
The first type of all-zero TC tiles is coming from batching subgraphs. Because there is no edge connection among nodes across different subgraphs. This type of all-zero TC tiles dominates the overall collected number of all-zero tiles.
The second type of all-zero tiles comes from the missing edge connections between the nodes within each subgraph. While this type of all-zero tiles is minor in its quantity compared with the first type. It potentially reduces memory access and computation. 

\begin{figure}[t] \small
    \centering
    \includegraphics[width=0.8\columnwidth]{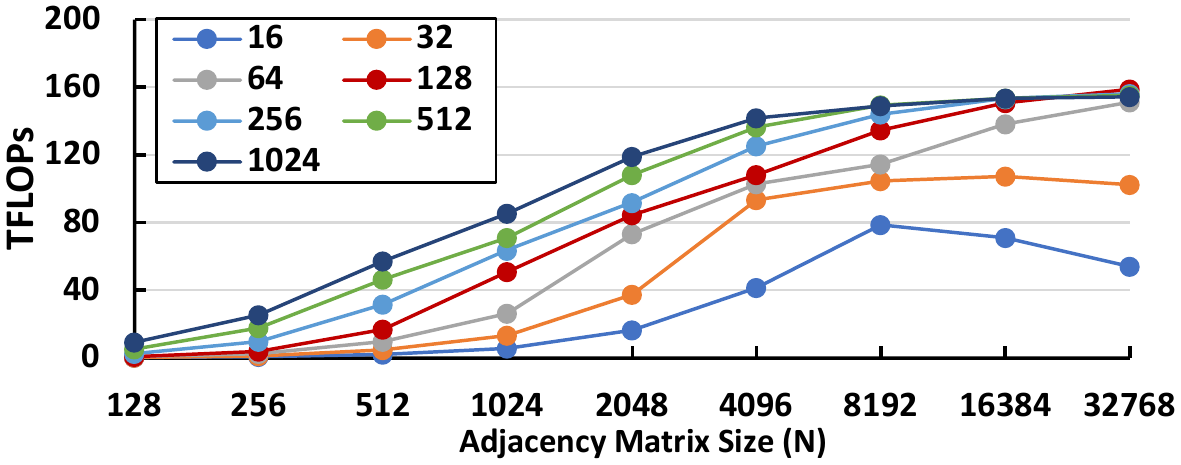}
    \vspace{-5pt}
    \caption{Adjacency matrix size impact. Note that we choose the common subgraph size N=\{128, 256, ..., 32768\} and the hidden embedding dimension D=\{16, 32, ..., 1024\}.}
    \label{fig: Adjacency matrix size impact.}
    \vspace{8pt}
\end{figure}

\vspace{2pt}
\textbf{Adjacency Matrix Size.}  
We will demonstrate the subgraph adjacency matrix size impact on the performance of \Mname. 
Specifically, adjacency matrix size can be controlled by specifying the \textit{number of subgraphs} (in METIS) and \textit{batching size} (in data loader). 
The size of the adjacency matrix will impact the performance of aggregation in terms of computations and data movements, {meanwhile, it will also determine whether our GNN computation can fully utilize the available GPU resources.}. 
We use 1-bit for both adjacency matrix and node embedding matrix in this study. 
As shown in Figure~\ref{fig: Adjacency matrix size impact.}, we can observe that under the same size of $D$, with the increase of the number of nodes ({i.e., the value of $N$}), our major 1-bit GEMM computation kernel would scale up its performance well. 
Note that different colored lines represent different embedding sizes, and we mainly focus on the computation of $\mathbf{AX}$ (\textit{i.e.}, $N\times N\times D$, where $N$ is the number of nodes and $D$ is the node embedding dimension) for neighbor aggregation phase. {in the settings of small subgraph size (128 to 512), the increase of the overall computation throughput is not evident, because the computation size is small and most of the available GPU resources such as SMs would achieve low utilization. While in the range of subgraph size (512 to 16,384), we can notice a more significant increase in the TFLOPs performance. Because in these settings, more computations from the bit-level data manipulation would trigger more SMs to participate in the BMM computation, thereby, improving the overall GPU throughput. For those large subgraph sizes (> 16,384) the overall throughput would hardly increase, mainly because all available GPU computation units are almost fully in use.}
One specialty of those batched GNN computations \textit{w.r.t.} the traditional NN computation is that batch GNN have more skewed-sized matrices in terms of the ratio between $N$ and $D$. 
This, to some degree, limits the achievable peak performance on TC. What is also worth noticing is that among different lines (different choices of $D$), the larger $D$ usually leads to better utilization of the GPU, since more computation and memory resources of the GPU will become active for higher throughput.

\hspace{2pt} 
\textbf{Non-zero Tile Reuse.}
We will demonstrate the effectiveness of our non-zero tile reuse by a control-variable study. We eliminate the number of non-zero tiles impact on performance by setting all tiles to non-zero tiles (\textit{i.e.}, filling the initial matrix with all ones). Then we choose the neighbor aggregation process ($\mathbf{\hat{X}}=\mathbf{A}\mathbf{{X}}$) for the study and fix the $D$ to 1024. We change $N$ from 1,024 to 8,192. Three bit combinations are used in our evaluation, where $\mathbf{A}$ is consistently using 1-bit while $\mathbf{X}$ is using 4, 8, and 16 bit. 
\begin{figure}[h] \small
    \centering
    \includegraphics[width=0.8\columnwidth]{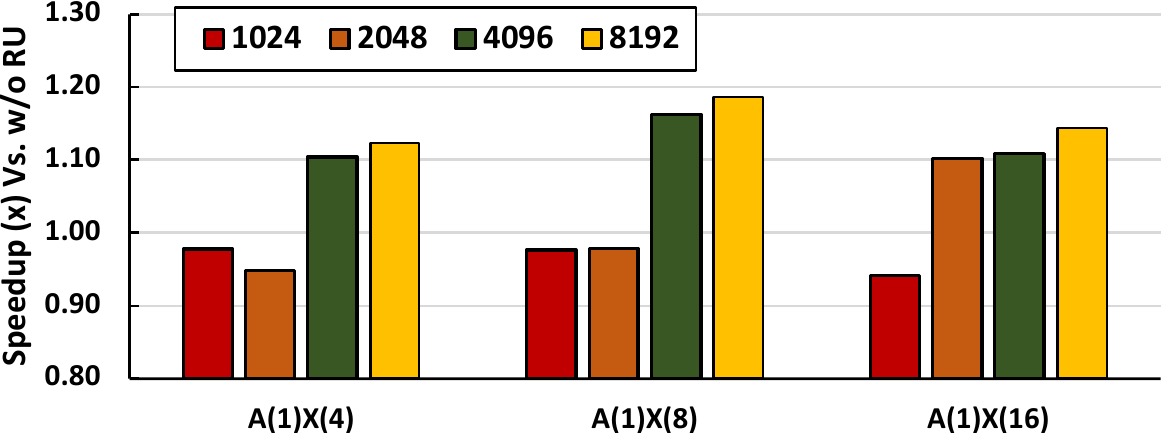}
    \caption{Non-zero tile reuse effectiveness. Note that we choose subgraph size N=\{1024,2048,4096,8192\} for this study.}
    \label{fig: Non-zero Tile Reuse Effectiveness.}
\end{figure} 

As described in Figure~\ref{fig: Non-zero Tile Reuse Effectiveness.}, our non-zero tile reuse can improve the throughput performance on those large matrix sizes with the higher number of bits. 
The major reason behind this is that reuse the non-zero tile can largely reduce the global memory access for fetching the same 1-bit adjacency matrix tile repetitively, which is the key performance bottleneck for those large metrics. 
The setting (w/o nonzero-tile) reuse shows more advantage on the smaller size matrix because the overhead of recurrent loading the same adjacency matrix tile is not pronounced compared with GEMM operations on TC. {This study inspires us to come up with a more intelligent strategy or heuristics to determine under which condition applying the non-zero tile reuse will bring performance benefits and we would leave this for our future work for exploration.}

\section{Conclusion}
In this paper, we propose \Mname, the first QGNN computing framework to support any-bitwidth computation via GPU Tensor Core.
Specifically, we introduce the first GNN-tailored any-bitwidth arithmetic design that can emulate different bitwidth computations to meet the end-users demands.
We craft a TC-tailored CUDA kernel design by incorporating 3D-stacked bit compression, zero-tile jumping, and non-zero tile reuse technique to maximize the performance gains from GPU Tensor Core.
We also incorporate an effective bandwidth-optimized subgraph packing strategy to maximize the data transferring efficiency.
Finally, we integrate QGTC with the popular PyTorch framework for better programmability and extensibility.
Extensive experiments show significant performance gains of QGTC in practice.

\section{Acknowledgment}
We would like to thank anonymous PPoPP paper reviewers for their valuable suggestions on our paper writing and PPoPP artifact reviewers for helping us improve our software artifact functionality and reusability to benefit future research. This work was supported by National Science Foundation under the award number 2124039. 
\bibliographystyle{ACM-Reference-Format}
\bibliography{reference}
\balance
\clearpage
\appendix
\section{Artifact Appendix}

\subsection*{Abstract Summary}
\Mname~is an efficient design and implementation for Quantized GNN computing on NVIDIA Ampere GPUs (e.g., A100 and RTX3090). 
\Mname~consists of two parts. 
The first part is the host-side CPU Python program. It is responsible for dataset loading, runtime configuration generation, and invoking the GPU-side program. 
The second part is the device-side GPU kernel. It is responsible for the major computation of the Quantized GNN model through floating-point number bit-decomposition and GEMM-based computation for quantized GNNs.
\Mname~improves the performance of Quantized GNN computing with its kernel design and optimization based on 1-bit Tensor Core primitive from NVIDIA Ampere Architecture. Moreover, the runtime configuration generation on the host-side CPU program makes \Mname~more adaptive towards various kinds of input settings.

\subsection*{Artifact Checklist}

\begin{itemize}
\itemsep0em 
    \item \textbf{Link:} \url{github.com:YukeWang96/PPoPP22_QGTC.git}.
    \item \textbf{Hardware}:
    \begin{itemize}
    \itemsep0em 
        \item Intel CPU x86\_64 with host memory >= 32GB. Tested on Intel Xeon Silver 4110 (8-core with 16-thread) CPU with 64GB host memory.
        \item  NVIDIA GPU (arch>=$sm\_80$) with devcie memory >= 16GB. Tested on NVIDIA RTX 3070 ($sm\_86$) and RTX3090 ($sm\_86$). Note that upon creating this artifact, we mainly evaluate our design on RTX3090. The execution time may be different across different devices but the overall speedup is similar.
    \end{itemize}
    \item \textbf{OS \& Compiler}:
    NVIDIA-Docker-2.0, Ubuntu 16.04+, GCC 7.5+, CMAKE 3.14+, CUDA 11.3.
\end{itemize}

\subsection*{Environment Setup}
\noindent \textbf{Step-1: Setup the basic environment.} Two options:
\begin{itemize}
\itemsep0em 
    \item Setup the environment via Docker (\textbf{Recommended}).
    \begin{itemize}
    \itemsep0em 
        \item Run \aecode{docker pull happy233/qgtc:updated}
        \item Run \aecode{docker run -it --rm --gpus all -v \$PWD/:/qgtc happy233/qgtc:updated /bin/bash}
    \end{itemize}
    \item Setup via \aecode{conda} and \aecode{pip}.
    \begin{itemize}
    \itemsep0em 
        \item Create a new conda environment: \aecode{conda create -n env\_name python=3.6}
        \item Activate conda environment: \aecode{conda activate env\_name}
        \item Install PyTorch: \aecode{conda install pytorch torchvision torchaudio cudatoolkit=11.1 -c pytorch -c conda-forge} and \aecode{pip install torch requests}.
        \item Install DGL: \aecode{conda install -c dglteam dgl-cuda11.0}.  
        \item Install QGTC: \aecode{TORCH\_CUDA\_ARCH\_LIST="8.6" python setup.py  clean --all install}
    \end{itemize}
\end{itemize}
Details of these options can be found in \aecode{README.md}.

\vspace{3pt}
\noindent \textbf{Step-2: Install \Mname~PyTorch Binding.}
\begin{itemize}
\itemsep0em 
    \item Go to \aecode{QGTC\_module/}
    \item Run \aecode{./build.sh} to install the QGTC modules for running QGTC kernel.
\end{itemize}

\noindent \textbf{Step-3: Download datasets.} 
We preprocess graph datasets in \aecode{.npy} format that can be downloaded and extracted automatically by running \aecode{./download\_dataset.sh}.
Note that node initial embedding is not included, and we generate an all 1s embedding matrix according to users input dimension parameter at the runtime for just performance evaluation.

\subsection*{Experiments}
\begin{itemize}
\itemsep0em 
    \item Running DGL baseline on GNN inference (Figure 7(a,b)).
    \begin{itemize}
    \itemsep0em 
        \item Go to the root directory of this project.
        \item Run \aecode{./1\_7a\_eval\_DGL\_cluster\_GCN.py} for the cluster GCN and \aecode{./1\_7b\_eval\_DGL\_batched\_GIN.py} for the batched GIN of the DGL baseline. Each script will automatically generate a \aecode{.csv} result file.
    \end{itemize}
    \item Running cuBLASgemmEX for INT8 GEMM kernel comparison (Figure 7(c)).
    \begin{itemize}
    \itemsep0em 
        \item Go to \aecode{cuBLASgemmEX/} directory.
        \item Run \aecode{./compile.sh} to compile cuBLAS baseline.
        \item Run \aecode{./bench\_cuBLAS\_INT8.py} to profile cuBLAS Tensorc Core GEMM in INT8 precision.
        \item Go to the project root directory.
        \item Run \aecode{./2\_7c\_QGTC\_GEMM\_INT8.py} to profile our \Mname~low-bit GEMM built on 1-bit Tensor Core primitive for comparison.
    \end{itemize}
    \item Running QGTC on the cluster GCN and the batched GIN (Figure 7(a,b)).
    \begin{itemize}
    \itemsep0em 
        \item Go to project root directory.
        \item Run \aecode{./0\_7a\_eval\_QGTC\_cluster\_GCN.py}  for the cluster GCN and \aecode{./0\_7b\_eval\_QGTC\_batched\_GIN.py} for the batched GIN and generate \aecode{.csv} result files.
    \end{itemize}
    \item Running some additional studies (Figure 8 and 9). Detailed commands of running all these studies can be found in \aecode{README.md}.
\end{itemize}

Note that in this artifact, we focus on the evaluation of the quantized GNN inference computation, and the reported time per epoch includes the quantized low-bit GNN model forward pass. We exclude the time of data loading and some other data preprocessing tasks.

\end{document}